\documentclass[preprint,showpacs,preprintnumbers,nofootinbib,amsmath,amssymb,aps,prc,longbibliography]{revtex4-1}
\usepackage{graphicx}
\usepackage{dcolumn}
\usepackage{bm}
\begin{document}
\title{Pairing reentrance in hot rotating nuclei}
\author{N. Quang Hung$^{1}$}
 \altaffiliation[On leave of absence from Center for Nuclear Physics, ]
 {Institute of Physics, Hanoi, Vietnam}
 \email{hung.nguyen@ttu.edu.vn}
\author{N. Dinh Dang$^{2,3}$}
  \email{dang@riken.jp}
 \affiliation{1) School of Engineering, TanTao University, TanTao University Avenue, TanDuc Ecity, Duc Hoa, Long An Province, Vietnam \\
2) Theoretical Nuclear Physics Laboratory, RIKEN Nishina Center
for Accelerator-Based Science,
2-1 Hirosawa, Wako City, 351-0198 Saitama, Japan\\
3) Institute for Nuclear Science and Technique, Hanoi, Vietnam}

\date{\today}
\begin{abstract}
The pairing gaps, heat capacities and level densities are calculated within the BCS-based quasiparticle approach including the effect of thermal fluctuations on the pairing field within the pairing model plus noncollective rotation along the $z$ axis for $^{60}$Ni and $^{72}$Ge nuclei. The analysis of the numerical results obtained shows that, in addition to the pairing gap, the heat capacity can also serve as a good observable to detect the appearance of the pairing reentrance in hot rotating nuclei, whereas such signature in the level density is rather weak.
\end{abstract}

\pacs{21.60.-n, 21.60.Jz, 24.60.-k, 24.10.Pa}
\keywords{Suggested keywords}
\maketitle
\section{Introduction}
\label{Intro}

In the collective rotation of a deformed nucleus, the
rotation axis is perpendicular to the symmetry axis, and the Coriolis 
force, which breaks the Cooper pairs, increases with the total angular momentum
so that at a certain critical angular momentum all Cooper pairs are broken.
The nucleus undergoes then a phase transition from the superfluid
phase to the normal one (the SN phase transition). This is
called the Mottelson-Valatin effect~\cite{Mottelson}. 
A similar effect is expected in spherical nuclei, 
although the rotation is no longer collective. The total
angular momentum is made up by those of the nucleons from the broken
pairs, which occupy the single-particle levels around the Fermi surface and block them
against the scattered pairs. The pairing correlations decrease until
a sufficiently large total angular momentum $M_{c}$, where the pairing gap
$\Delta$ completely vanishes. At finite temperature ($T\neq$ 0), the
increase of $T$ relaxes the tight packing of
quasiparticles around the Fermi surface, which is caused by a large
angular momentum $M\geq M_{c}$, and spreads them farther away from the
Fermi level. This makes some levels become partially unoccupied,
therefore, available for scattered pairs. 
As the result, when $T$ increases up to some 
critical value $T_{1}$, the pairing correlations are energetically
favored, and the pairing gap reappears. As $T$ goes higher, the 
increase of a large number of quasiparticles eventually breaks down the
pairing gap at $T_{2}$ $(>T_{1})$. This phenomenon, predicted by
Kamuri~\cite{Kammuri} and Moretto~\cite{Moretto}, is called thermally
assisted pairing or anomalous pairing, and  
later as pairing reentrance by Balian, Flocard and
V$\Acute{\rm e}$n$\Acute{\rm e}$roni~\cite{Balian}. 

However, it has been shown already in the 1960s that the sharp SN-phase transition at $M=M_{c}$ in
the Mottelson-Valatin effect is an artifact of the BCS method. As a
matter of fact, a proper particle-number 
projection before variation has removed 
the discontinuity in the pairing gap as a decreasing function of
the angular momentum~\cite{Mang}. 
Similarly, by taking the effect of thermal fluctuations 
in the pairing field into account, the SN phase transition
predicted by the BCS theory is smoothed out. The gap $\Delta(T)$ of a non-rotating
nucleus does not collapse at 
$T_{c}\simeq 0.568\Delta(T=0)$, but monotonically decreases with
increasing $T$, remaining finite even at $T\gg
T_{c}$~\cite{Moretto2,MBCS}. This result is reconfirmed by shell-model
calculations of pairing energy as a function of excitation
energy~\cite{Zele}, and
by embedding the exact eigenvalues of the pairing problem into the canonical
ensemble~\cite{Exact}. By considering an 
exactly solvable cranked deformed shell model Hamiltonian
it has also been shown that the pairing gap, 
quenched at $T =$ 0 and high rotational 
frequency, reappears at $T_{1}$ ($\ll T_{c}$) at $M\geq M_{c}$~\cite{Frauendorf}.
However, different from the prediction by 
the BCS theory, the pairing gap does not vanish at $T>T_{1}$.

The behavior of hot rotating nuclei can be put in correspondence with superconductors in the presence of an external
magnetic field, where the magnetic field plays the role as that of the
nuclear rotation. The reentrance of superconducting correlations, which is also known as the unconventional superconductivity, has been the subject of recent theoretical and experimental studies in condensed matter. The Grenoble High Magnetic Field Laboratory has recently discovered that URhGe becomes superconducting at low temperature in the presence of a strong magnetic field (between 8 and 13 T), well above the field value of 2 T, at which superconductivity is first destroyed~~\cite{Sci}. The reentrance of superconductivity under the magnetic field is interpreted as to be caused by spin reorientation~\cite{reent}, which bears some similarity with the reappearance of the scattered pairs in rotating hot nuclei discussed above.

Recently, we have developed an approach 
based on the finite-temperature BCS (FTBCS) that includes 
the effects due to quasiparticle-number fluctuations in the
pairing field and the $z$ projection of angular momentum at $T\neq$ 0, which we call 
as the FTBCS1 (with "1" denoting the effect due to quasiparticle-number fluctuations)~\cite{SCQRPA}. 
This approach reproduces well the effect 
of smoothing out the SN phase transition at $T\neq$ 0 
as well as the pairing reentrance in hot (noncollectively) 
rotating nuclei. In the latter case, for $M\geq M_{c}$ 
the pairing gap also reappears at $T_{1}$ and 
remains finite at $T>T_{1}$. 
It has also been pointed out that the microscopic
mechanism of the nonvanishing gap at high $T$ is the quasiparticle-number fluctuations, which are 
ignored in the conventional BCS theory. A refined version of the
FTBCS1 also includes the contribution of coupling to pair vibrations within 
the self-consistent quasiparticle random-phase approximation~\cite{SCQRPA}.

That the pairing reentrance is not an artifact of the mean field analysis, but a robust physical effect, has been obtained by exact diagonalization of the 2D attractive Hubbard model, where a nonmonotonic filed dependence of the pair susceptibility in the presence of the external magnetic field was found for various cluster sizes both in the weak and strong coupling limit~\cite{Gorczyca}. 
Nonetheless, the experimental extraction of the pairing gap in hot nuclei
is not simple because one has to properly exclude the admixture with the
contribution of uncorrelated single-particle configurations 
from the odd-even mass difference~\cite{Ensemble}. 
Therefore the detection of the
pairing reentrance effect by using the experimentally extracted
pairing gaps seems to be elusive, 
especially when the formally derived pairing gap has a value smaller than the average spacing between the single-particle levels. 

Meanwhile, 
the heat capacity has been extracted from the experimental 
level densities~\cite{EXP}. The existence of a bump or an $S$ shape
on the curve of the heat capacity
at $T\sim T_{c}$ allows one to discuss about the smoothing of the SN phase
transition in finite nuclei. In a recent calculation of 
the heat capacity in $^{72}$Ge within the Shell Model Monte
Carlo (SMMC) approach, by reconfirming the 
pairing reentrance effect, the authors of Ref. \cite{Dean} claimed 
that they found a local dip in the heat capacity at 
rotation frequency of 0.5 MeV at $T\sim$ 0.45 MeV, and a corresponding local maximum on the
temperature dependence of the logarithm of level density.
They associated such irregularities in the heat capacity and level density as the signatures of the pairing reentrance. 
There are, however, two concerns regarding these results.
The first one is that, as well-known, at such low temperature the
SMMC approach produces quite large
error bars. As a consequence, instead of approaching zero as 
it should be to fulfill the 
third law of thermodynamics, the SMMC heat capacity at $T<$ 0.5 MeV jumps 
to 25 $\sim$ 30 (Fig. 4 of \cite{Dean}), which makes the statement on 
the signature of the pairing reentrance ambiguous.
The second one is that the SMMC in Ref.  \cite{Dean} used the same Fock-space 
single-particle energies of the shells ($0f1p - 0g1d2s$)  for both neutron and proton spectra. 
Hence, the difference between neutron and proton spectra 
came solely from the difference in the valence particle numbers 
outside the closed-shell core of $^{40}$Ca, which ignored altogether 
the Coulomb barrier in the proton spectrum. 
Our FTBCS1 theory is free from such deficiency at low
$T$, and it works well with the schematic 
as well as single-particle spectra, which are obtained from the realistic 
Woods-Saxon potential. Therefore, in the present paper we will calculate the 
heat capacity as well as the level density within the FTBCS1 theory to see if these quantities can be used to identify the pairing reentrance phenomenon in realistic nuclei at finite temperature and angular momentum. 

The paper is organized as follows. The formalism for the calculations of thermodynamic quantities such as pairing gap, heat capacity, and level density in hot non-collectively rotating nuclei within the FTBCS and FTBCS1 theories  is presented in Sec. \ref{formalism}.  The results of numerical calculations are analyzed in Sec. \ref{results}. The paper is summarized in the last section, where conclusions are drawn.
\section{Formalism}
\label{formalism}
We consider the pairing Hamiltonian describing a spherical system 
rotating about the symmetry $z$ axis \cite{Moretto}:
\begin{equation}
H = H_{P}-\lambda\hat{N}-\gamma\hat{M}~,  \label{Hamil}
\end{equation}
where $\lambda$ and $\gamma$ are the chemical potential and rotation frequency, respectively. $H_{P}$ is the standard pairing Hamiltonian of a system, which consists $N$ particles
interacting via a monopole pairing force with the constant parameter $G$ (the
BCS pairing Hamiltonian), namely
\begin{equation}
H_P=\sum_k\epsilon_k(a_{+k}^{\dagger}a_{+k}+a_{-k}^{\dagger}a_{-k}) - G\sum_{kk'}{a_k^{\dagger}
a_{-k}^{\dagger}a_{-k'}a_{k'}}~, \label{Hpair}
\end{equation}
with $a_{\pm k}^{\dagger}(a_{\pm k})$ being the creation (annihilation) operators of a particle (neutron or proton) with angular momentum $k$, projection $\pm m_k$, and energy $\epsilon_k$. The particle-number operator $\hat{N}$ and total angular momentum $\hat{M}$, which coincides with its $z$ projection, are given as
\begin{equation}
\hat{N}=\sum_k(a_{+k}^{\dagger}a_{+k} + a_{-k}^{\dagger}a_{-k})~, \hspace{5mm}
\hat{M}=\sum_k m_k(a_{+k}^{\dagger}a_{+k} - a_{-k}^{\dagger}a_{-k}) ~. \label{NM}
\end{equation}
After the Bogoliubov transformation from the particle operators, $a_k^\dagger$ and  $a_k$, to the quasiparticle ones, $\alpha_k^\dagger$ and $\alpha_k$,
\begin{equation}
a_k^\dagger = u_k\alpha_k^\dagger + v_k\alpha_{-k}~, \hspace{5mm}
a_{-k}=u_k\alpha_{-k}  - v_k\alpha_k^\dagger ~, \label{Bogo}
\end{equation}
the Hamiltonian \eqref{Hamil} is transformed into the quasiparticle one $\cal H$, whose explicit form can be found, e.g., in Refs. \cite{SCQRPA,SCQRPA1}.

As has been discussed in Refs. \cite{Kammuri,Moretto,SCQRPA}, for a spherically symmetric system, the laboratory-frame $z$ axis, which is taken as the axis of quantization, can always be made coincide with the body-fixed one, which is aligned with the direction of the total angular momentum within the quantum mechanical uncertainty. Therefore the total angular momentum is completely determined by its $z$-projection $M$ alone. For systems of an axially symmetric oblate shape rotating about the symmetry axis, which in this case is the principal body-fixed one, this noncollective motion is known as ``single-particle" rotation. The pairing reentrance effect was originally obtained within the BCS theory in 
Refs. \cite{Kammuri,Moretto} by considering such systems described by Hamiltonian (\ref{Hamil}). Its physical interpretation based on the thermal effect, which relaxes the tight packing of
quasiparticles around the Fermi surface due to a large
angle momentum $M\geq M_{c}$, and spreads them farther away from the
Fermi level, fits well in the framework of this ``single-particle" rotation. However, as has been pointed out in Ref. \cite{Moretto}, for non-spherical nuclei, and specifically in the case of axially symmetric ones, the spin and angular momentum projections on the symmetry axis are not good quantum numbers. In this case the formalism used here is not completed because it does not include the angular momentum's  component perpendicular to the symmetry axis. Cranking model might serve as a better solution of the problem in this situation. This remains to be investigated because the results obtained within the Lipkin model with $J_x$ cranking did not reveal any pairing reentrance so far~\cite{Civi}. On the other hand, in the region of high level densities (at high excitation energies and/or high $T$) the values of angular momentum projection on the symmetry axis will be mixed among the levels, which worsen the axial symmetry. The melting of shell structure will also eventually drive nuclei to their average spherical shape.
\subsection{FTBCS1 equations at finite angular momentum}
\label{gapeq}

The FTBCS1 includes a set of FTBCS-based equations, corrected by the effects of quasiparticle-number fluctuations, for
the level-dependent pairing gap $\Delta_{k}$, average particle number $N$, and average angular momentum $M$. The derivation of the FTBCS1 equations was reported in detail in Ref. \cite{SCQRPA}, so we do present here only the final equations. The FTBCS1 equation for the pairing gap is written as a sum of two parts, the level-independent part $\Delta$ and the level-dependent part $\delta\Delta_k$, namely
\begin{equation}
\Delta_k = \Delta + \delta\Delta_k ~, \label{Gapeq}
\end{equation}
where 
\begin{equation}
\Delta=G\sum_{k'}{u_{k'}v_{k'}(1-n_{k'}^+-n_{k'}^-)}~, \hspace{5mm}
\delta\Delta_k=G\frac{\delta{\cal N}_k^2}{1-n_k^+-n_k^-}u_kv_k ~, \label{Gapeq1} 
\end{equation}
where
\begin{eqnarray}
u_k^2&=&\frac{1}{2}\left(1+\frac{\epsilon_k-Gv_k^2-\lambda}{E_k}\right)~, \hspace{5mm}
v_k^2=\frac{1}{2}\left(1-\frac{\epsilon_k-Gv_k^2-\lambda}{E_k}\right)~, \nonumber \\
E_k&=&\sqrt{(\epsilon_k-Gv_k^2-\lambda)^2+\Delta_k^2}~, \hspace{5mm}
n_k^{\pm} = \frac{1}{1+e^{\beta(E_k \mp \gamma m_k)}}~,\hspace{5mm} \beta=1/T~. \label{uv}
\end{eqnarray}
with the quasiparticle-number fluctuations $\delta{\cal N}_k^2$ at nonzero angular momentum
\begin{equation}
\delta{\cal N}_k^2=(\delta{\cal N}_k^+)^2+(\delta{\cal N}_k^-)^2 = n_k^+(1-n_k^+)+n_k^-(1-n_k^-) ~. \label{QNF}
\end{equation}
The corrections due to coupling to pair vibration beyond the quasiparticle mean field at finite temperature and angular momentum are significant only in light nuclei like oxygen or neon isotopes, whereas they are negligible for medium and heavy nuclei (See Figs. 6 - 8 of Ref. \cite{SCQRPA}).
As the present paper considers two medium nuclei, $^{60}$Ni and $^{72}$Ge, these corrections on the FTBCS1 equations are neglected in the numerical calculations.  

The equations for the particle number and total angular momentum are the same as Eqs. (25) and (26) of Ref. \cite{SCQRPA}, namely
\begin{equation}
N = 2\sum_k\left[v_k^2(1-n_k^+ -n_k^{-}) + \frac{1}{2}(n_k^\dagger+n_k^{-}) \right], \hspace{5mm}
M = \sum_k m_k(n_k^+ - n_k^{-}) ~. \label{NM1}
\end{equation}
The system of coupled equations \eqref{Gapeq} -- \eqref{NM1} are called the FTBCS1 equations at finite angular momentum. Once the FTBCS1 equations are solved, the total energy ${\cal E}$, heat capacity $C$ and entropy $S$ of the system are calculated
\begin{eqnarray}
{\cal E} &=& \langle{\cal H}\rangle~, \hspace{5mm} C = \frac{\partial{\cal E}}{\partial T}~, \nonumber \\
S&=& -\sum_k[n_k^+ {\rm ln} n_k^+ + (1-n_k^+){\rm ln}(1-n_k^+)+ n_k^{-} {\rm ln} n_k^{-} + (1-n_k^{-}){\rm ln}(1-n_k^{-})] ~. \label{ECS}
\end{eqnarray}

\subsection{Level density}
\label{NLD}

Within the conventional FTBCS, the level density is calculated as the invert Laplace transformation of the grand partition function \cite{Moretto}
\begin{equation}
\rho(E, N, M)=\frac{1}{(2\pi i)^3}\oint d\beta \oint d\alpha \oint d\mu e^S ~, \label{Rho}
\end{equation}
where $\alpha=\beta\lambda$, $\mu =\beta\gamma$, and $S$ is entropy of the system. The saddle-point approximation gives a good evaluation of the integral \eqref{Rho}. As the result, the total level density of a system with $N$ neutrons and $Z$ protons is given as 
\begin{equation}
\rho(E,N,M) = \frac{e^S}{(2\pi)^2\sqrt{D}} ~, \label{Rho1}
\end{equation}
where $S = S_N + S_Z$ and 
\begin{equation}
D = \left| \begin{array}{cccc}
\frac{\partial^2\Omega}{\partial\alpha_N^2} & \frac{\partial^2\Omega}{\partial\alpha_N\partial\alpha_Z} & \frac{\partial^2\Omega}{\partial\alpha_N\partial\mu} & \frac{\partial^2\Omega}{\partial\alpha_N\partial\beta} \\
\frac{\partial^2\Omega}{\partial\alpha_Z\partial\alpha_N} & \frac{\partial^2\Omega}{\partial\alpha_Z^2} & \frac{\partial^2\Omega}{\partial\alpha_Z\partial\mu} & \frac{\partial^2\Omega}{\partial\alpha_Z\partial\beta} \\
\frac{\partial^2\Omega}{\partial\mu\partial\alpha_N} & \frac{\partial^2\Omega}{\partial\mu\partial\alpha_Z} & \frac{\partial^2\Omega}{\partial\mu^2} & \frac{\partial^2\Omega}{\partial\mu\partial\beta} \\
\frac{\partial^2\Omega}{\partial\beta\partial\alpha_N} & \frac{\partial^2\Omega}{\partial\beta\partial\alpha_Z} & \frac{\partial^2\Omega}{\partial\beta\partial\mu} & \frac{\partial^2\Omega}{\partial\beta^2}
\end{array} \right | ~. \label{Det}
\end{equation}
The logarithm of the grand-partition function of the systems is given as
\begin{equation}
\Omega = \Omega_N + \Omega_Z = S + \alpha_N N +\alpha_Z Z + \mu M - \beta{\cal E}~. \label{Omega}
\end{equation}
The derivation of $\Omega$ with respect to $\alpha$ and $\mu$ can be seen explicitly in Eqs. (25)-(35) of Ref. \cite{Moretto}. Within the FTBCS1, the grand-partition function has the same form as that given by Eq. \eqref{Omega} of the FTBCS. Therefore, the first and second derivatives of the FTBCS1 grand-partition function are the same as those of the FTBCS ones. The only difference  comes from the first derivatives of the pairing gap with respect to $\alpha$, 
$\mu$, and $\beta$ because of the quasiparticle-number fluctuations in the FTBCS1 gap equation \eqref{Gapeq1}. In this case, instead of  the simple Eqs. [(33)-(35)] of Ref. \cite{Moretto}, the derivatives become rather complicate expressions, which are obtained by taking the first derivatives of the left and right-hand sides of Eq. \eqref{Gapeq1} with respect to $\alpha$, $\mu$, and $\beta$. Amongst the three derivatives of the FTBCS1 gap, ${\partial\Delta_{k}}/{\partial\beta}$ can be obtained by using its definition, namely
\begin{equation}
\frac{\partial\Delta_{k}}{\partial\beta}=-T^{2}\frac{\partial\Delta_{k}}{\partial T}=-T^{2}\frac{\Delta_{k}(T+\delta T)-\Delta_k(T)}{\delta T} ~, 
\label{Dgap}
\end{equation}
and can be easily calculated numerically by choosing an appropriate value of $\delta T$ as the input parameter. The other two derivatives, ${\partial\Delta_{k}}/{\partial\alpha}$ and 
${\partial\Delta_{k}}/{\partial\mu}$, must be calculated from their explicit analytic expressions because $\alpha$ and $\beta$ are two Lagrange multipliers, which are obtained by solving the FTBCS1 equations. 
The final equations for ${\partial\Delta_k}/{\partial\alpha}$ and ${\partial\Delta_k}/{\partial\mu}$ are derived as
\begin{equation}
\sum_k\left(A_k\beta\frac{\partial\Delta_k}{\partial\alpha}+B_k \right)+\left(C_k^+ + C_k^- -\frac{2}{G}\right)\beta\frac{\partial\Delta_k}{\partial\alpha}+(D_k^+ + D_k^-) = 0~, \label{Dgap1}
\end{equation}
\begin{equation}
\sum_k\left(A_k'\beta\frac{\partial\Delta_k}{\partial\mu}+B_k' \right)+\left(C_k' + D_k' -\frac{2}{G}\right)\beta\frac{\partial\Delta_k}{\partial\mu}+E_k' = 0 ~, \label{Dgap2}
\end{equation}
where
\begin{equation}
A_k=\frac{1}{E_k^3}\left\{(\lambda-\epsilon_k)^2(1-n_k^+-n_k^-)+\beta\Delta_k^2E_k^2[n_k^++n_k^--2(n_k^+)^2-2(n_k^-)^2] \right\} ~, 
\label{1}
\end{equation}
\begin{equation}
B_k=-\frac{\Delta_k(\lambda-\epsilon_k)}{E_k^3}\left\{1-n_k^+-n_k^- -\beta E_k^2(n_k^++n_k^- -2(n_k^+)^2-2(n_k^-)^2 \right\} ~, 
\label{2}
\end{equation}
\begin{equation}
C_k^+=\frac{n_k^+(1-n_k^+)}{(1-n_k^+-n_k^-)^2E_k^3}\left\{(\lambda-\epsilon_k)^2(1-n_k^+-n_k^-) - \beta\Delta_k^2 E_k[(n_k^+ -1)^2+n_k^-(2n_k^+ -n_k^-)] \right\} ~, 
\label{3}
\end{equation}
\begin{equation}
C_k^-=\frac{n_k^-(1-n_k^-)}{(1-n_k^+ -n_k^-)^2E_k^3}\left\{(\lambda-\epsilon_k)^2(1-n_k^+-n_k^-) - \beta\Delta_k^2 E_k[(n_k^- -1)^2+n_k^+(2n_k^- -n_k^+)] \right\} ~, 
\label{4}
\end{equation}
\begin{equation}
D_k^+=-\frac{n_k^+(1-n_k^+)\Delta_k(\lambda-\epsilon_k)}{(1-n_k^+-n_k^-)^2E_k^3}\{1-n_k^+ -n_k^- +\beta E_k[(n_k^+ -1)^2 +n_k^-(2n_k^+ -n_k^-)] \} ~, 
\label{5}
\end{equation}
\begin{equation}
D_k^-=-\frac{n_k^-(1-n_k^-)\Delta_k(\lambda-\epsilon_k)}{(1-n_k^+ -n_k^-)^2E_k^3}\{1-n_k^+ -n_k^- +\beta E_k[(n_k^- -1)^2 +n_k^+(2n_k^- -n_k^+)] \} ~, 
\label{6}
\end{equation}
\begin{equation}
A_k'=\frac{1}{E_k^3}\{(\lambda-\epsilon_k)^2(1-n_k^+-n_k^-)+\beta\Delta_k^2E_k[n_k^+(1-n_k^+) +n_k^-(1-n_k^-)] \} ~, 
\label{7}
\end{equation}
\begin{equation}
B_k'=\frac{\beta m_k\Delta_k}{E_k}[n_k^+(1-n_k^+) + n_k^-(1-n_k^-)] ~, 
\label{8}
\end{equation}
\begin{equation}
C_k'=\frac{(\lambda-\epsilon_k)^2}{(1-n_k^+-n_k^-)E_k^3}[n_k^+(1-n_k^+) + n_k^-(1-n_k^-)] ~,
\label{9}
\end{equation}
\[
D_k'=-\frac{\beta\Delta_k^2}{(1-n_k^+-n_k^-)^2E_k^2}\{n_k^+(1-n_k^+)[(n_k^+-1)^2+n_k^-(2n_k^+-n_k^-)]
\]
\begin{equation}
+ n_k^-(1-n_k^-)[(n_k^--1)^2+n_k^+(2n_k^--n_k^+)] \} ~, 
\label{10}
\end{equation}
\begin{equation}
E_k'=-\frac{\beta m_k\Delta_k}{E_k}[n_k^+(1-n_k^+)+n_k^-(1-n_k^-)] ~.
\label{11}
\end{equation}
By solving first the FTBCS1 equations, then Eqs. (\ref{Dgap1}) and (\ref{Dgap2}), one obtains ${\partial\Delta_k}/{\partial\alpha}$ and ${\partial\Delta_k}/{\partial\alpha}$ as functions of $T$ at a given value of the total angular momentum $M$. 
\section{Analysis of numerical results}
\label{results}
The numerical calculations are carried out for two realistic $^{60}$Ni and $^{72}$Ge nuclei. The latter is considered in order to have a comparison with the results obtained within the SMMC approach in Ref. \cite{Dean}. 
The single-particle spectra for these two nuclei are obtained within the axially deformed Woods-Saxon potential \cite{WS},
whose parameters are chosen to be the same as those given in Ref. \cite{leveldens}. All the bound (negative energy) single-particle states are used in the calculations. The quadrupole deformation parameters $\beta_2$ are equal to 0 and -0.224 for $^{60}$Ni and 
$^{72}$Ge, respectively. The pairing interaction parameters are adjusted so that the pairing gaps at $T=0$ fit the experimental values obtained from the odd-even mass differences. These values are $G_N =$ 0.347 MeV, which gives $\Delta_N=$ 1.7 MeV for neutrons in $^{60}$Ni and $G_N =$ 0.291 MeV, which gives $\Delta_N=$ 1.7 MeV for neutrons in $^{72}$Ge. For protons  in $^{72}$Ge, the value $G_Z =$ 0.34 MeV is chosen to give $\Delta_Z=$ 1.5 MeV, whereas there is no pairing gap for the closed-shell protons ($Z=$ 28) in $^{60}$Ni.
Because the pairing gap (\ref{Gapeq}) is level-dependent, the level-weighted gap $\bar{\Delta}$ is considered, which is defined as
$\bar{\Delta}=\sum_k\Delta_k/\Omega$ with the total number $\Omega$ of levels in the deformed basis [In the case of spherical basis it becomes $\bar{\Delta} = \sum_j(2j+1)\Delta_j/\sum_j(2j+1)$~\cite{SCQRPA}].

    \begin{figure}
       \includegraphics[width=14cm]{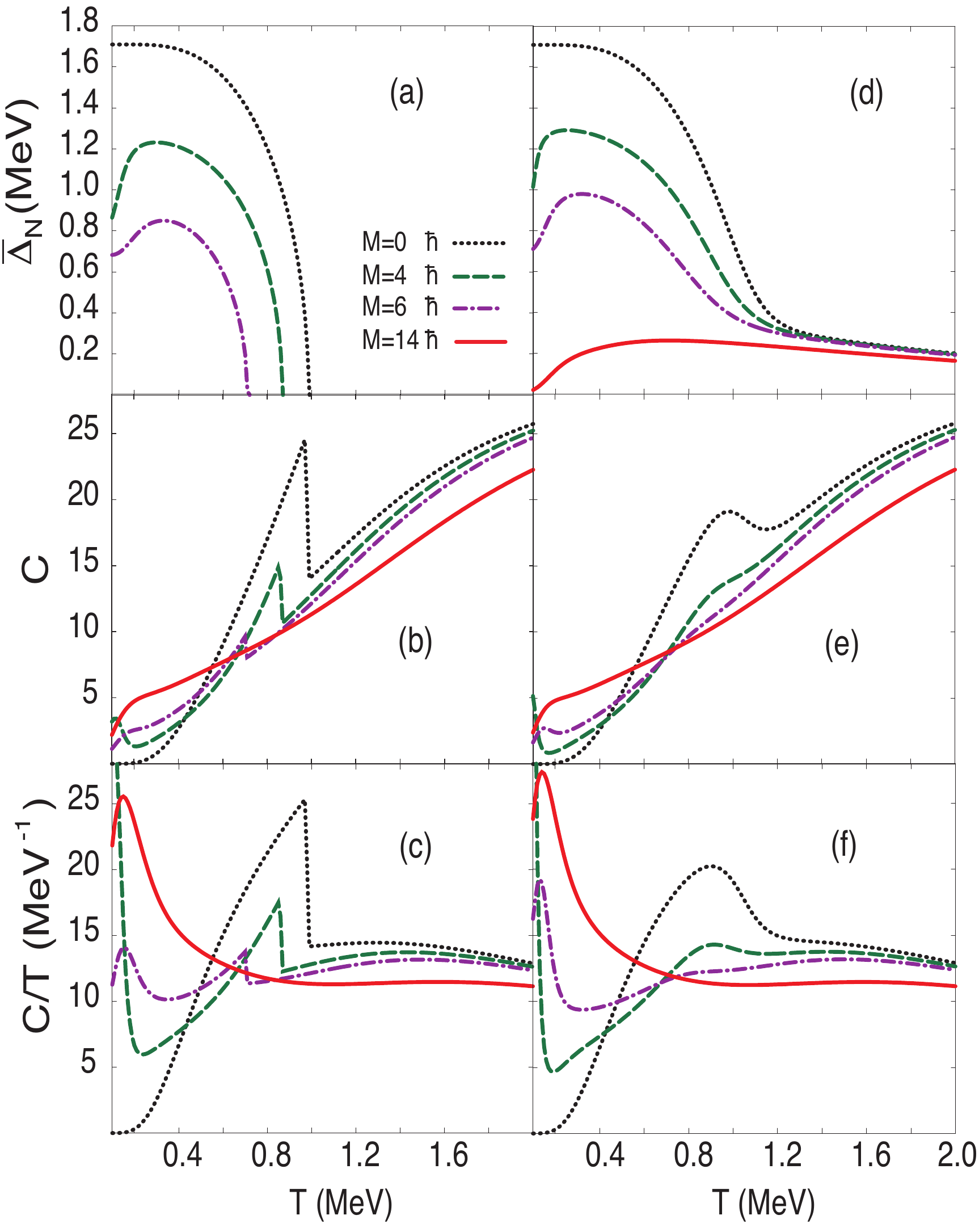}
        \caption{(Color online)  Level-weighted neutron pairing gap $\bar{\Delta}$ [(a), (d)], heat capacity $C$ 
        [(b), (e)], and heat capacity divided by 
        temperature $C/T$ [(c), (f)] for $^{60}$Ni obtained at different values of angular momentum $M$ 
       as functions of $T$. Panels (a) - (c) show the FTBCS results, whereas the predictions by the FTBCS1 are displayed in
       (d) - (f).
        \label{gapNi60}}
    \end{figure}
Shown in Fig. \ref{gapNi60} are the neutron level-weighted pairing gap $\bar{\Delta}$, 
the heat capacity $C$, and the ratio $C/T$ obtained as functions of $T$ at several values of the total angular momentum $M$.
The left column represents the predictions by the standard FTBCS, whereas the results obtained within the FTBCS1 are displayed in the right column. Both approaches show the pairing reentrance in the gaps at $M=$ 4 and 6 $\hbar$, namely the gap increases with $T$ up to $T\simeq$ 0.3 MeV, then decreases as $T$ increases further. Because of the quasiparticle-number fluctuations, the FTBCS1 gap does not collapse at $T_c$ as the FTBCS one, but decreases monotonically at high $T$. At $M=$ 14 $\hbar$, while the FTBCS gap completely vanishes at all $T$, the FTBCS1 gap shows a spectacular reentrance effect, namely it increases from the zero value at $T=$ 0 up to around 0.3 MeV at $T\simeq$ 0.8 MeV, and then slowly decreases as $T$ further increases.

The heat capacities obtained within the FTBCS and FTBCS1 look alike, except for the region around $T_c$, where the quasiparticle-number fluctuations smooth out the sharp SN phase transition so that the sharp local maximum is depleted to a broad bump. In the region, where the pairing reentrance takes place, namely at $T\simeq$ 0.3 MeV and $M=$ 4 or 6 $\hbar$, a weak local minimum is seen on the curve representing the temperature dependence of the heat capacity similarly to the feature reported in Ref. \cite{Dean}.  This local minimum is magnified by using the ratio $C/T$ so that the latter might be useful in experiments as a quantity to identify the pairing reentrance. However, when the gap is too small as in the pairing reentrance at $M=$ 14 $\hbar$,  the heat capacity $C$ ($C/T$) obtained within FTBCS1 is almost identical to that predicted by the FTBCS, where the gap is zero.

    \begin{figure}
       \includegraphics[width=12cm]{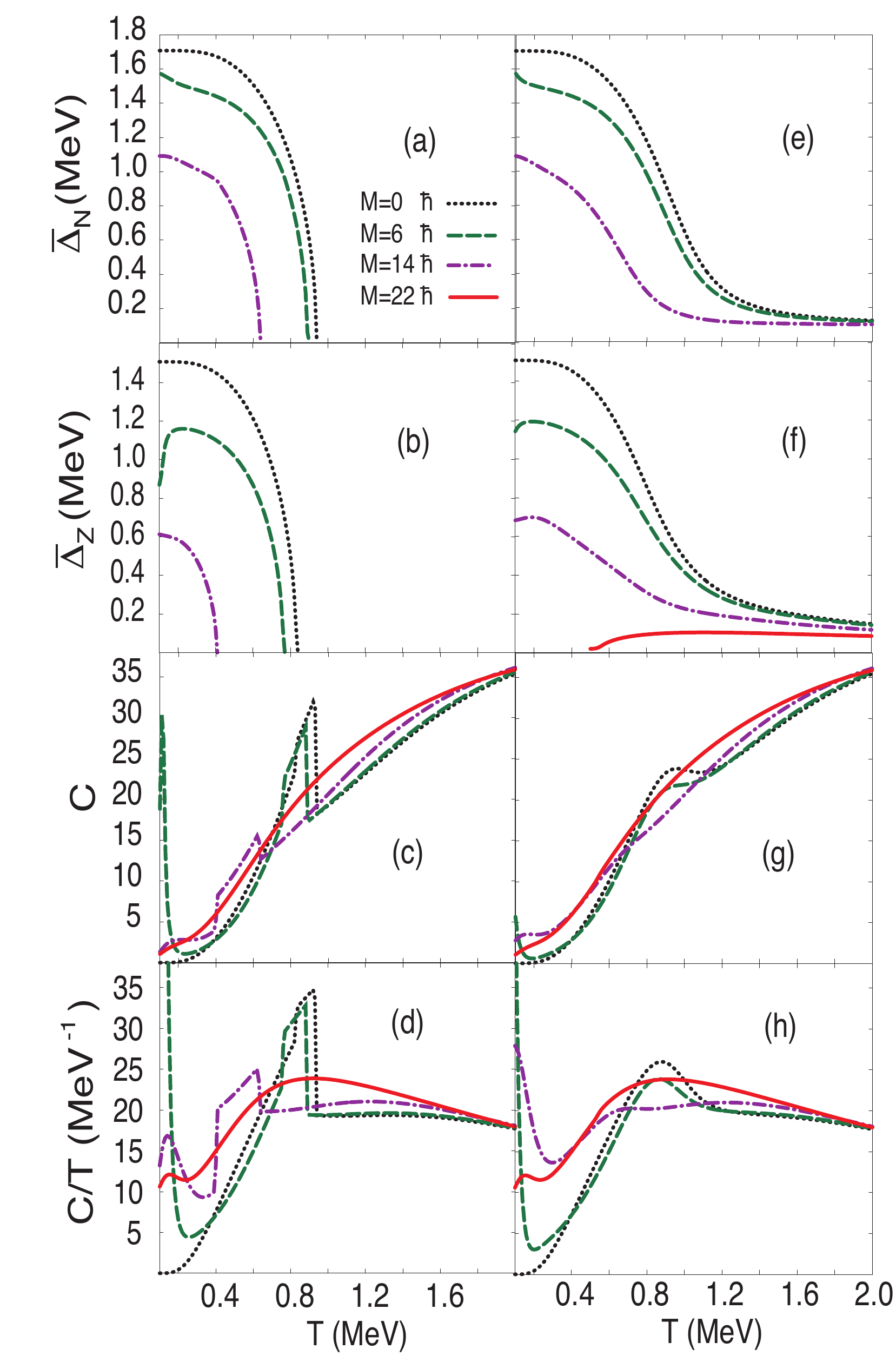}
        \caption{(Color online)  Level-weighted pairing gaps for neutrons [(a), (e)], protons [(b), (f)], heat capacity $C$ 
        [(c), (g)], and heat capacity divided by 
        temperature $C/T$ [(d), (h)] for $^{72}$Ge obtained at different values of angular momentum $M$ 
       as functions of $T$. Panels (a) - (d) show the FTBCS results, whereas the predictions by the FTBCS1 are displayed in
       (e) - (h).
        \label{gapGe72}}
    \end{figure}
For $^{72}$Ge, both neutron and proton gaps exist, which cause two peaks in the temperature dependence of the heat capacity obtained within the FTBCS, as shown in Fig. \ref{gapGe72} (c). The overall features of $C$ and $C/T$ for $^{72}$Ge are
similar to those obtained for $^{60}$Ni. As compared with the results of Ref. \cite{Dean}, where the same single-particle energies in the ($0f1p - 0g1d2s$) shells were used for both neutrons and protons, and where the pairing reentrance was predicted for neutrons, no pairing reentrance effect for neutrons is seen in the results of our calculations. On the other hand, the pairing reentrance takes place
for protons at $M\geq$ 6 $\hbar$, as shown in Fig. \ref{gapGe72} (b) and \ref{gapGe72} (f).

    \begin{figure}
       \includegraphics[width=16cm]{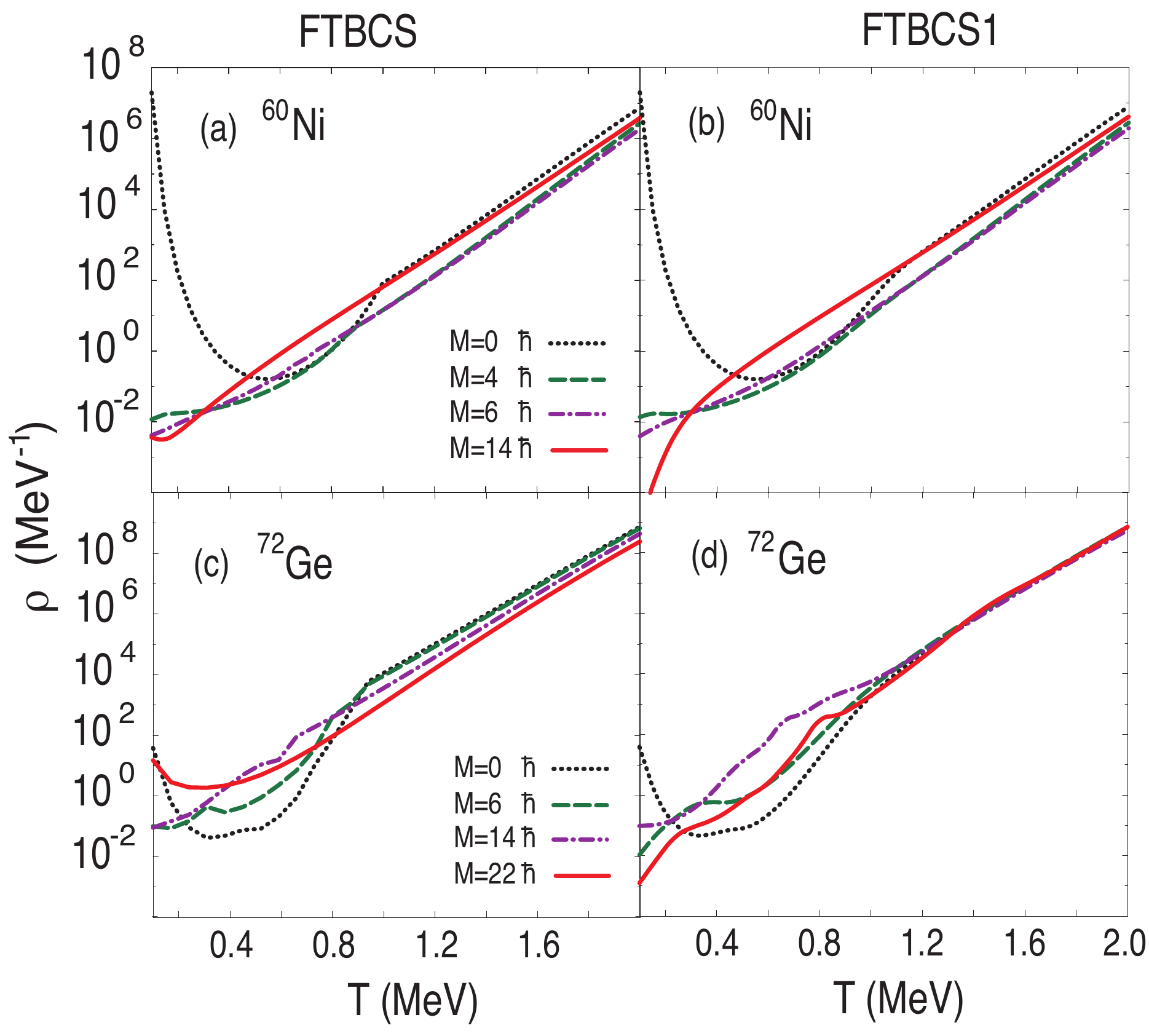}
        \caption{(Color online)  Level densities as functions of $T$ at several values of total angular momentum $M$
        obtained for $^{60}$Ni [(a), (b)] and $^{72}$Ge [(c), (d)] within the FTBCS (left panels) and FTBCS1 (right panels).
                \label{rhoNiGe}}
    \end{figure}
Another experimentally measurable quantity, which may help to identify the pairing reentrance effect, is the level density. In fact, the authors of Ref. \cite{Dean} claimed that the pairing reentrance causes an irregularity in a shape of a small local maximum 
at low $T$ on the curve, which describes the temperature dependence of the level density. The level densities obtained at several values of the total angular momentum $M$ for $^{60}$Ni and $^{72}$Ge are displayed in Fig. \ref{rhoNiGe} as functions of $T$. These results show a trend of transition of the level density from  a convex function of $T$ to a concave function after the pairing reentrance occurs. This is particularly clear for $^{60}$Ni by comparing the FTBCS1 predictions for the level density at $M<$ 14 $\hbar$,  which are convex functions of $T$, with that obtained at $M=$ 14 $\hbar$, which is a concave function of $T$. For $^{72}$Ge this trend is less obvious because of the existence of proton and neutron pairing gaps with different values of $T_c$
within the FTBCS. However, contrary to the result shown in the inset of Fig. 4 in Ref. \cite{Dean}, no pronounced local maximum that might correspond to the pairing reentrance is seen here in the temperature dependence of the level density for $^{72}$Ge. Since the results for the pairing gap, heat capacity, and level density strongly depend on the selected single-particle energies, the irregularity seen in the temperature dependence of the level density at $\omega=$ 0.5 MeV in the inset of Fig. 4 of Ref. \cite{Dean} might well be an artifact caused by 
using the same single-particle energies for both neutron and proton spectra. 

    \begin{figure}
       \includegraphics[width=16cm]{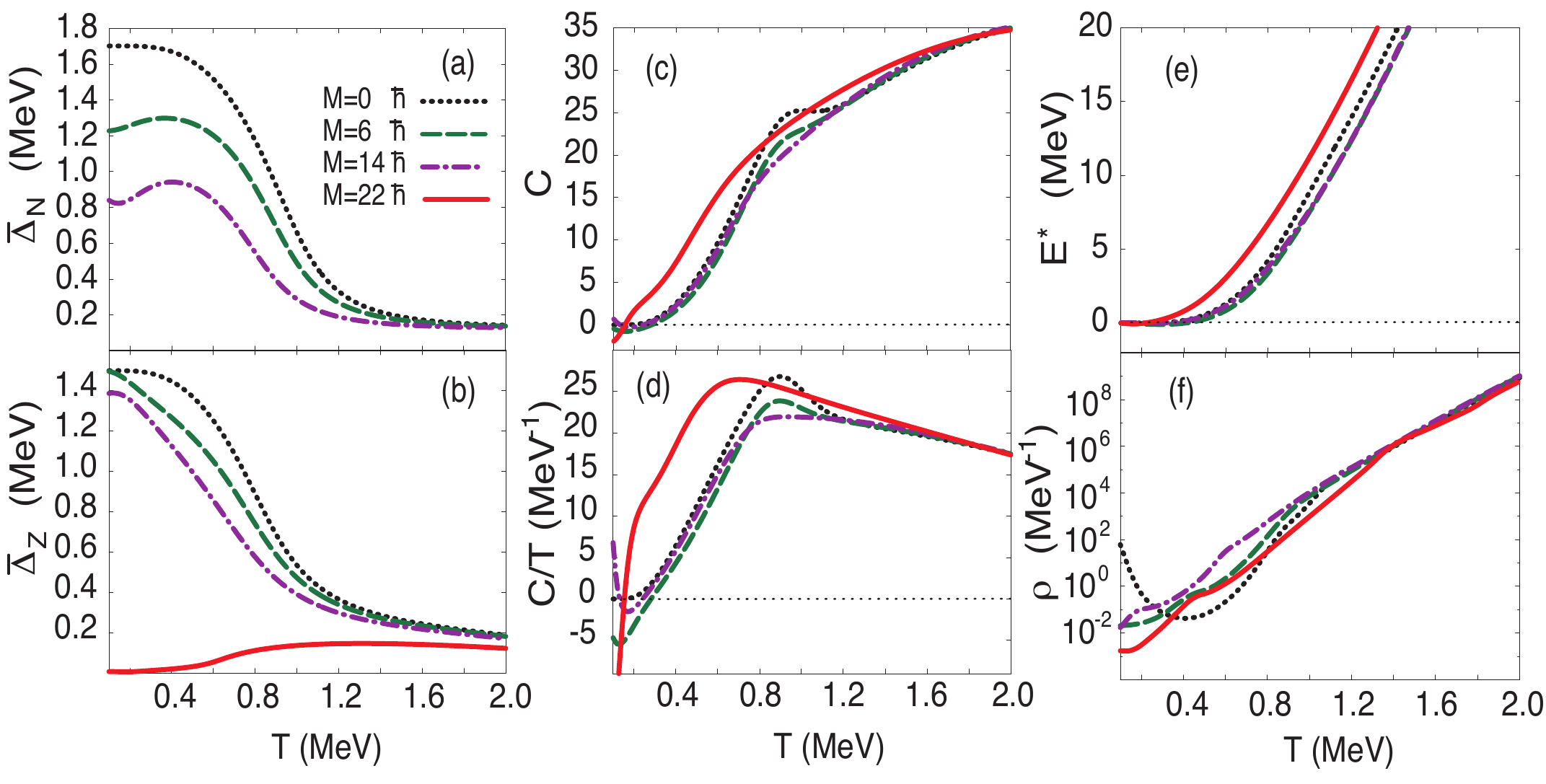}
        \caption{(Color online)  Level averaged neutron and proton gaps $\bar\Delta_{N,Z}$, heat capacity $C$, $C/T$, 
        excitation energy $E^{*}$ and level density for $^{72}$Ge as functions of $T$ at several values of $M$ [shown in (a)] obtained within the FTBCS1 in the test calculations by using the shells $(0f1p - 0g1d2s)$ atop the $^{40}$Ca core with the values of Woods-Saxon neutron single-particle energies adopted for both neutron and proton spectra.
                        \label{testGe}}
    \end{figure}
To show that it is indeed the case, we carried out the test calculations by using only the $(0f1p - 0g1d2s)$ shells on top of the $^{40}$Ca core with the values of Woods-Saxon neutron single-particle energies  adopted for both neutron and proton spectra of $^{72}$Ge, as in Ref. \cite{Dean}. The difference now solely comes from that between the numbers of valence neutrons and protons (20 valence neutrons and 12 valence protons). The results of these test calculations within the FTBCS1 are shown in Fig. \ref{testGe}. They show the pairing reentrance in the neutron pairing gap instead of the proton one, at low $M$. This is in qualitative agreement with the pairing reentrance predicted for neutrons in Fig. 2 of Ref. \cite{Dean}, where no proton pairing reentrance is seen up to $\omega=$ 0.5 MeV. In our calculations, however, the proton pairing reentrance takes place at rather high $M=$ 22 $\hbar$. In either case, the pairing reentrance is so strong that causes the excitation energy $E^{*}$ to decrease slightly with increasing $T$ at low $T$. This violates the second law of thermodynamics. As a consequence, at $M\geq$ 6 $\hbar$, the heat capacity becomes negative at $T<$ 0.4 MeV. The results for the level density remain essentially the same as compared to those previously obtained by using all proton and neutron single-particle levels, but with only bound-state single particle energies [Fig. \ref{rhoNiGe}  (d)], and no irregularities such as a pronounced local maximum
are found. In our opinion, the pick on the dotted curve in the inset of Fig. 4 in Ref. \cite{Dean} emerges because of the two lower values of ${\rm ln}\rho$ at $T\simeq$ 0.42 and 0.45 MeV. These two lower values are the results obtained by calculating the level density in the canonical ensemble  $\rho(E) = \beta e^{S}/\sqrt{2\pi C}$, making use of the two large values of $C$ equal to around 16 and 28 (with large error bars). However, these large values of the heat capacity at low $T$ are the artifacts of the SMMC calculations because the heat capacity must be zero (or very small) at $T=$ 0 (or very low $T$) to avoid an infinite (or very large) entropy, which would violate the third law of thermodynamics. Therefore, we conclude that the neutron pairing reentrance effect, reported in Ref. \cite{Dean}, is caused by the use of the same single-particle spectrum for both protons and neutrons, whereas the irregularity seen on the curve of ${\rm ln}\rho$ in the inset of Fig. 4 of Ref. \cite{Dean} is caused by unphysically large values of the heat capacity at low $T$ in the SMMC technique.

The present calculations within the FTBCS and FTBCS1 do not take into account the effects of residual interactions beyond the 
monopole pairing one. It is well known that these effects are responsible for strong collective motion in finite nuclei, which leads to the increase of nuclear level density.  In spherical nuclei the collective enhancement of level density is caused by vibrational excitations, whereas in deformed nuclei it comes from the collective rotation. The contribution of collective motion to the increase of nuclear level density has been studied in detail by Ignatyuk and collaborators starting from the early 1970s~\cite{Ignatyuk,Junghans}. Because the Hamiltonian used in the SMMC calculations of Ref. \cite{Dean} included the quadrupole-quadrupole interaction, let us estimate the effect of the collective quadrupole vibration on the increase of level density. By using the adiabatic approximation (3) of Ref. \cite{Junghans} for the enhancement coefficient $K_{vib}$ due to quadrupole vibration, and the experimental energies $E(2^{+}_{1})$ of the lowest quadrupole excitation  in Ref. \cite{1st2}, we found $K_{vib}\simeq$ 1.06 and 1.94 for $^{60}$Ni at $T=$ 0.3 MeV and $^{72}$Ge at $T=$ 0.4 MeV, respectively. These values of T are those, at which the pairing reentrance starts to show up in these nuclei.  
With the deformation parameter $\beta_2=$ -0.224 adopted in the present calculations for $^{72}$Ge
 and by using Eq. (9) of Ref. \cite{Zagreb}, we found the enhancement coefficient $K_{coll}(\beta_2)\simeq$ 1.96 for $^{72}$Ge at $T=$ 0.4 MeV, whereas for the spherical nucleus, $^{60}$Ni, $K_{coll} = K_{vib}$ = 1.06 at $T=$ 0.3 MeV.  Therefore, for both nuclei, 
$^{60}$Ni and $^{72}$Ge, one can expect that the collective quadrupole enhancement of level density is not dramatic at the value of temperature, where the pairing reentrance is supposed to take place.  The contribution of collective motion generated by higher multipolarities to the increase of level density is expected to be much smaller.
In Ref. \cite{combi} the quasiparticle Tamm-Dancoff Approximation, which includes the isoscalar quadrupole-quadrupole interaction and $J_{x}$ cranking, was used to calculate the level density within the microcanonical ensemble. The authors of Ref. \cite{combi} found 4 $\leq K_{rot}\leq$ 6 and 1.002 $\leq K_{vib}\leq$ 1.012 at excitation energy 3 $\leq E\leq$ 8 MeV (i.e. at around 0.38 $\leq T\leq$ 0.63 MeV) for $^{162}$Dy. They also found a monotonic decrease of the average pairing gap with increasing the excitation energy up to $E=$ 8.5 MeV ($T\simeq$ 0.65 MeV), i.e. much higher than $T_c \simeq$ 0.34 MeV. These results are in good qualitative agreement with our estimations.

Finally, it is worth noticing that the inclusion of the approximate particle-number projection within the Lipkin-Nogami method does not significantly alter the behavior of the paring reentrance obtained within the FTBCS1 theory (See Fig. 6 of Ref. \cite{SCQRPA}). This does not diminish the value of an approach based on exact particle-number and angular momentum projections. In Ref. \cite{Horoi} the exact solution of the nuclear shell model is used to study the SN phase transition including residual interactions other than the pairing one. The results of Ref. \cite{Horoi}, which fully respect the particle number and angular momentum conservations, confirm the presence of a long tail of pair correlations far beyond the BCS phase transition region in agreement with the prediction by the FTBCS1. The approach of Ref. \cite{Horoi} does not use any external heat bath, which determines the temperature of thermal equilibrium. Therefore the nuclear temperature can only be extracted from the level density by using the Clausius definition of thermodynamic entropy. This task is not easy because of the discrete and finite nuclear spectra (See, e.g. Ref. \cite{Ensemble} and references therein). Nonetheless, instead of using temperature, it would be interesting to see if the pairing reentrance takes place in the pair correlator as a function of excitation energy at various values of angular momentum within the method of Ref. \cite{Horoi}.
\section{Conclusions}
The present paper studies the temperature dependences of the heat capacity and level density in hot medium-mass nuclei, which undergo a noncollective rotation about the symmetry axis.  The numerical calculations, carried out by using the realistic Woods-Saxon single-particle energies for $^{60}$Ni and $^{72}$Ge within the FTBCS and FTBCS1 theories, have shown the pairing reentrance in the pairing gap at finite angular momentum $M$ and temperature $T$. Instead of decreasing with increasing $T$, the gap first increases with $T$ then decreases at higher $T$.  It is demonstrated that the heat capacity $C$, or rather $C/T$, and level density $\rho$ can be used to experimentally identify the pairing reentrance effect. The pairing reentrance, when it occurs, leads to a clear depletion in the temperature dependence of the heat capacity, whereas the level density weakly changes from a convex function of $T$ to a concave one. 

Regarding the appearance of the local minimum in the heat capacity because of the pairing reentrance, the results of the present paper agree with that of the SMMC calculations in Ref. \cite{Dean}. However, the present results show no pronounced local maximum in the temperature dependence of the level density. The pairing reentrance is seen in the proton pairing gap of $^{72}$Ge at low $M$, whereas Ref. \cite{Dean} reported this effect in the neutron pairing energy. The test calculations by using the same single-particle configuration as that used in Ref. \cite{Dean}, but obtained within the Woods-Saxon potential, reveals that  the neutron pairing reentrance in $^{72}$Ge is an artifact, which is caused by the use of the same single-particle spectrum for both protons and neutrons, whereas the irregularity on the curve for the logarithm of level density, reported in Ref. \cite{Dean}, is caused by unphysically large values of the heat capacity at low $T$ in the SMMC approach. 
\acknowledgments
The numerical calculations were carried out using the FORTRAN IMSL
Library by Visual Numerics on the RIKEN Integrated Cluster of Clusters (RICC) system. NQH acknowledges the support by the National Foundation for Science and Technology Development(NAFOSTED) of Vietnam through Grant No. 103.04-2010.02. He also thanks the Theoretical Nuclear Physics Laboratory of RIKEN Nishina Center for its hospitality during his visit in RIKEN.

\end{document}